\documentclass[]{spie}  
\usepackage[]{graphicx}

\def\farcs{\hbox{\kern 0.13ex.\kern -0.95ex%
\raisebox{.9ex}{\scriptsize$\prime\prime$}\kern -0.1ex}}

\title{Recent Results and Perspectives for Precision Astrometry and Photometry 
with Adaptive Optics.} 

\author{
Jessica R. Lu\supit{a}, 
Andrea M. Ghez\supit{b,c},
Sylvana Yelda\supit{b}, 
Tuan Do\supit{b},
Will Clarkson\supit{b},
Nate McCrady\supit{d},
Mark Morris\supit{b}
\skiplinehalf
\supit{a}California Institute of Technology, MC 249-17, Pasadena, CA 91125; \\
\supit{b}UCLA Department of Physics and Astronomy, Los Angeles, CA 90095; \\
\supit{c}UCLA Institute of Geophysics and Planetary Physics, Los Angeles, CA 90095; \\
\supit{d}University of Montana, Missoula, MT 59812; 
}

\authorinfo{Send correspondence to jlu@astro.caltech.edu}

\begin{document} 
  \maketitle 

\begin{abstract}
Large ground-based telescopes equipped with adaptive optics (AO) systems 
have ushered in a new era of high-resolution infrared photometry and astrometry.
Relative astrometric accuracies of $<$0.2 mas have already been 
demonstrated from infrared images with spatial resolutions of 55-95 mas 
resolution over 
10-20'' fields of view. Relative photometric accuracies of 3\% and absolute 
photometric accuracies of 5\%-20\% are also possible. I will review 
improvements and current limitations in astrometry and photometry with
adaptive optics of crowded stellar fields. 
These capabilities enable experiments such as measuring 
orbits for brown dwarfs and exoplanets, studying our Galaxy's 
supermassive black hole and its environment,
and identifying individual stars in young star clusters, which can
be used test the universality of the initial mass function.
\end{abstract}


\section{INTRODUCTION}
\label{sec:intro}  

Scientific results from adaptive optics (AO) systems on ground-based telescopes 
have increased dramatically over the last $\sim$10 years. 
As existing AO systems have matured and stabilized
and the number of scientifically productive AO systems has increased,
AO observations have become more routine. With this stability, we
are now exploring the most effective ways to extract high-precision
quantitative science from AO observations. In this proceeding, we review
the current state of astrometric (\S2) and photometric (\S3 and \S4) accuracies 
achieved with existing AO systems. We also highlight the limitations 
of current measurements in an effort to determine where improvements
can be made. Each section begins
with a list of several science results utilizing AO
astrometric or photometric measurements and the accuracies achieved, followed
by a discussion of the limiting factors. Precision AO astrometry, 
in particular, is behind some of the driving science cases for future
AO systems and large telescopes (e.g. TMT, E-ELT). 
Understanding the current and
future limitations of AO astrometric and photometric measurements is still a
work in progress.


\section{Relative Astrometry}
The increased spatial resolution provided by adaptive optics observations 
in the near-infrared not only enables observations of crowded fields 
or close binary stars, but also provides improved astrometric precision. 
AO astrometry has been used to measure dynamical masses of brown dwarfs
\cite{dupuy09,konopacky10}, 
study our Galaxy's central supermassive black hole and its environment
\cite{ghez08,gillessen09,lu09,bartko09},
and study the formation and evolution of young star clusters
\cite{stolte08} and massive stars \cite{cameronThesis}. 

\begin{figure}
\begin{center}
\includegraphics[height=4in]{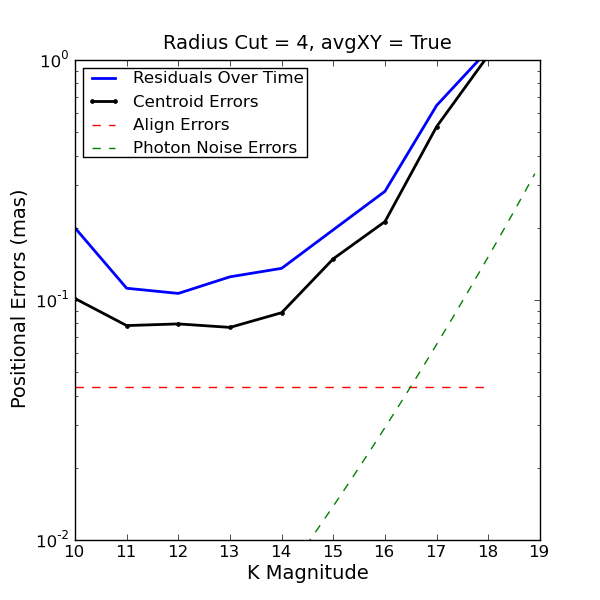}
\end{center}
\caption{
Astrometric error as a function of stellar brightness for Keck/NIRC2
LGS AO observations of stars in the Galactic center. 
The average random error for each data set, hereafter referred to as the
''centroiding error'' ({\em black solid}), is determined
by running the Starfinder PSF fitting routine on 3 subsets of the data 
and calculating the error on the mean position for each star. 
Each star's motion over time is fit with a linear motion model and the
residuals from these fits give the astrometric accuracy ({\em blue solid}).
Errors due to aligning data sets are negligible ({\em red dashed}).
Errors due to photon noise are calculated by simulating star fields with 
input positions and fluxes and a known PSF then comparing to the fluxes
recovered by Starfinder ({\em green dashed}). 
\label{fig:astrometryErr}
}
\end{figure}

Astrometric accuracies of $<$0.2 mas have been demonstrated
with the Keck and Palomar AO systems over multiple nights, months, 
and even years of observing \cite{cameron09,luThesis,clarkson10}. 
These accuracies are best determined by
measuring the residuals from linear motion for many stars within a
crowded-field, such as the Galactic center (see Figure \ref{fig:astrometryErr},
blue line). 

\begin{figure}
\begin{center}
\includegraphics[height=4in]{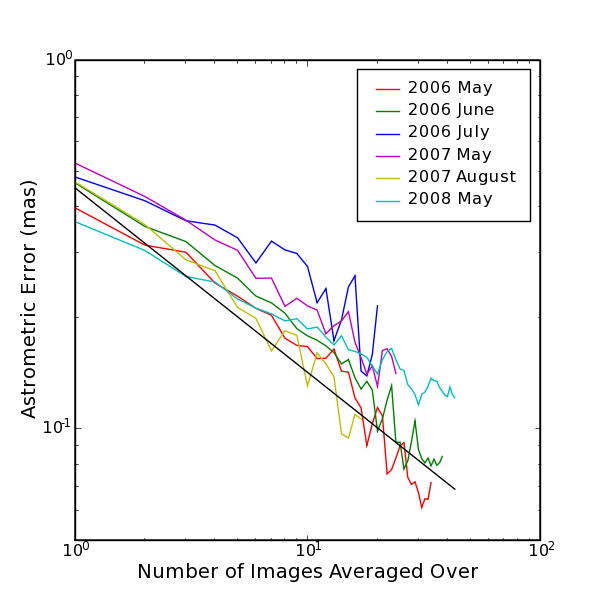}
\end{center}
\caption{
Reproduced from Lu\cite{luThesis}, figure 4.56. 
Astrometric error vs. integration time for several Keck LGS AO data sets
of the Galactic center. The data closely follow the expected trend of 
$1/\sqrt{t_{int}}$ shown by the {\em solid black line}.
\label{fig:allanVariance}
}
\end{figure}

Astrometric observations from ground-based telescopes are fundamentally
limited by differential tip-tilt jitter, which, fortunately, reduces
with integration time as $1/\sqrt{t_{int}}$ \cite{teoste88}. This behavior 
has been verified on sky by plotting the astrometric precision as a function of
total integration time for AO observations taken within a single night
(Figure \ref{fig:allanVariance}) \cite{cameron09,luThesis}. 
Practically,
astrometric measurements are more often limited by systematic effects
due to residual optical distortions, differential atmospheric refraction
\cite{gublerTytler98},
and PSF errors. Both optical distortions from the camera and differential
atmospheric refraction (DAR) can be calibrated and corrected. For instance,
Yelda et al. (in preparation) have used on-sky observations of globular
clusters to characterize the Keck NIRC2 distortion 
solution to the 1 mas level, a factor of 3 improvement over
previous solutions using pinhole masks. The relative distortion error felt
between epochs is also sensitive to the observational design.
To achieve the highest astrometric precision, residual distortions are 
minimized in Keck/UCLA Galactic center astrometric measurements
by always observing with small dithers\footnote{We still dither in order
to remove the effects of bad pixels, reduce the impact of intra-pixel
sensitivity variations, and average over small-scale distortions that
can't easily be characterized.} 
and at the same detector orientation 
and position
\cite{ghez08}.
Without this careful observing planning, residual distortions may be the 
dominant source of astrometric error for bright, uncrowded sources (see
Figure 14 in Fritz et al.\cite{fritz10}).
Uncorrected differential atmospheric refraction can also contribute 1-3 mas
of error for two stars separated by 5''; however,
it can be modeled and images corrected using real-time weather data and
a simple model of the Earth's atmosphere.

\begin{figure}
\begin{center}
\includegraphics[height=4in]{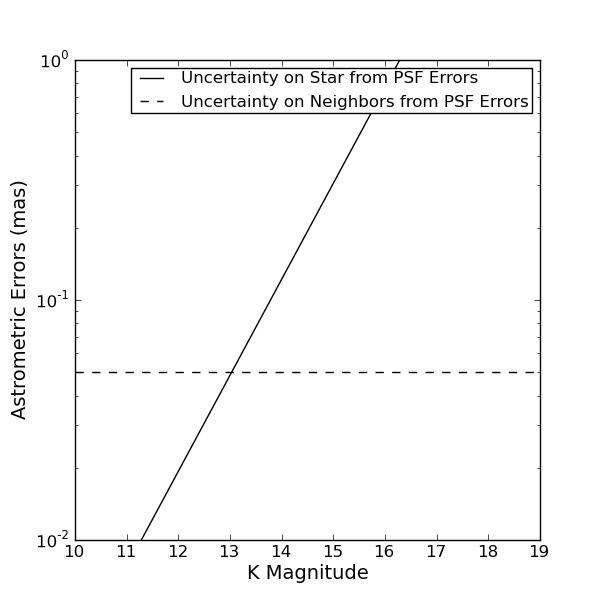}
\end{center}
\caption{
Astrometric errors vs. stellar brightness resulting from
PSF errors for VLT NAOS/CONICA observations
of the Galactic center \cite{fritz10}. At the bright end, their astrometry is
limited by residual distortion (see Figure 14 in Fritz et al. \cite{fritz10})
since their data sets are widely dithered. However, after this, the dominant
source of astrometric error at the bright end is from errors in the PSF
fitting process when using the wrong PSF to extract a star, a 
brightness-independent process.
At the faint end, astrometry is also limited by errors in the PSF
(see Equation 10 in Fritz et al. \cite{fritz10}).
This source of error is a result of incorrectly subtracted PSFs from 
brighter neighboring stars resulting in residuals that bias the astrometry of 
fainter sources; hence the error depends on stellar brightness. 
\label{fig:fritzAstrometry}
}
\end{figure}

The limiting factor for crowded-field AO astrometry appears to be PSF errors,
as shown by 
Fritz et al.\cite{fritz10}, even when there are plenty of stars to derive an 
empirical PSF using codes such as Starfinder \cite{starfinder}. 
Figure \ref{fig:fritzAstrometry}
shows astrometric errors vs. stellar brightness resulting from
PSF errors for VLT NAOS/CONICA observations of the Galactic 
center as described by Fritz et al. \cite{fritz10}. 
At the bright end, their astrometry is
limited by residual distortion (see Figure 14 in Fritz et al. \cite{fritz10})
since their data sets are widely dithered. However, after this, the dominant
source of astrometric error at the bright end is from errors in the PSF
fitting process when using the wrong PSF to extract a star, a 
brightness-independent process.
At the faint end, astrometry is also limited by errors in the PSF
(see Equation 10 in Fritz et al. \cite{fritz10}).
This source of error is a result of incorrectly subtracted PSFs from 
brighter neighboring stars resulting in residuals that bias the astrometry of 
fainter sources; hence the error depends on stellar brightness. 
These PSF errors depend on the Strehl of the observations, the amount of
PSF variability over the field of view, and the exact crowding of the 
science field being observed, so these numbers are not exactly applicable
to other observations. However, PSF errors are still likely to be a significant 
source of error for most crowded-field astrometric experiments.
In summary, calculations by Clarkson et al. (in preparation) suggest that 
the total unaccounted for error between the accuracies calculated from 
stellar motion residuals and the centroiding errors is $\sim$0.15 mas 
for bright sources in
Keck NIRC2 observations of the crowded Galactic center Arches cluster
at several epochs.
When combined with the centroiding error within each epoch, this results 
in an astrometric accuracy of 0.2 mas.

\section{Relative Photometry}

Relative photometry from adaptive optics observations has been used
to study both the time variability of individual sources and
the flux ratios or luminosity functions of multiple systems and
clusters of stars.
For instance, AO observations of Sgr A*-IR, the accreting supermassive black 
hole at the Galactic center, have been used to study its time variability
within a single night \cite{hornstein07,eckart08,do09}).
Also, AO observations have been used to construct luminosity functions
in massive young star clusters, which can be converted into 
initial mass functions in order to probe how star formation may differ
in the most crowded star clusters \cite{harayama08}.

\begin{figure}
\begin{center}
\includegraphics[height=4in]{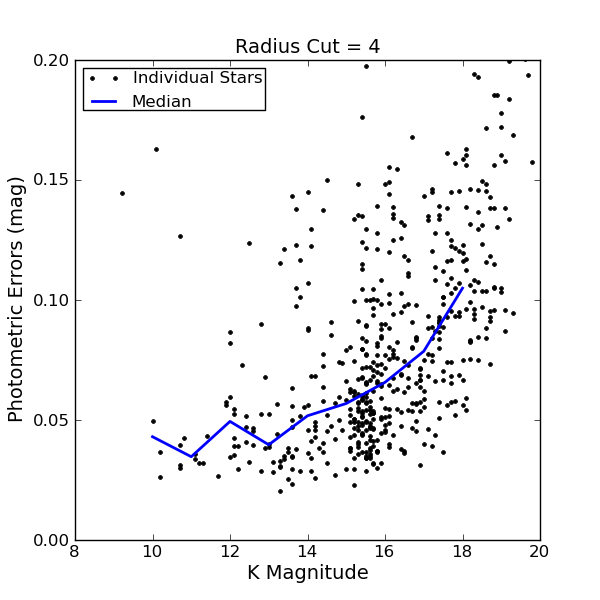}
\end{center}
\caption{
Relative photometric errors for AO observations of the Galactic center
at K'-band as a function of stellar brightness. The photometric errors
are calculated as the RMS error of 11 individual photometric measurements
taken over several years. For bright stars, the median photometric
accuracy is $\sim$4\%. No cuts have been made to remove intrinsically 
variable sources, which may contribute to the large spread in photometric
errors. The individual data sets were photometrically calibrated using
10 non-variable stars. The data sets are taken with Keck/NIRC2 LGS AO 
and are described in detail in
Yelda et al. (in preparation) and references therein.
\label{fig:relPhotometry}}
\end{figure}

Relative photometric accuracies of 3-4\% have been demonstrated with
AO observations. These accuracies can be reached within a single 
night of AO observations, under the same atmospheric and AO conditions,
as has been demonstrated \cite{doThesis,schoedel10} from RMS photometric
errors for individual stars across many exposures throughout a night.
Additionally, by comparing photometry for crowded star fields such 
as the Galactic center across many nights taken over several years, 
we show in Figure \ref{fig:relPhotometry} that the RMS photometric
errors of 4\% are still reached. 
Current limitations on relative photometry are due to errors
in the PSF estimates in these crowded fields. Analysis by
Schoedel et al.\cite{schoedel10} shows that extracting the PSF from images
using different PSF stars fully accounts for the relative photometric 
errors and is orders of magnitude above the photon noise limit for bright
stars (also see Esslinger \& Edmunds \cite{esslingerEdmunds98}).

\section{Absolute Photometry}
\label{sec:absPhotometry}

Absolute photometry is necessary for multi-wavelength analysis such
as for studies of stellar populations \cite{harayama08,stolte10}.
Typically, AO observations are absolutely photometrically calibrated
by using existing seeing-limited or HST observations taken in different
filters and with very different resolutions, requiring filter conversions
and crowding analysis. Absolute photometry of new fields typically
requires observations
of standard stars over a range of airmasses on a photometric night.
We have attempted such a program using the Keck AO system
and the NIRC2 instrument, observing 3 standard stars in natural guide star
(NGS) mode at airmasses of 1.0-1.8 in H, K', and L' filters. Using standard IRAF
aperture photometry reduction packages and a 2'' aperture radius,
we measure the zeropoints for these
three filters with uncertainties of 0.02 mag, 0.02 mag, and 0.06 mag at
H, K', and L', respectively. However, applying these zeropoints to
crowded-field LGS AO observations, where the size of the empirically 
estimated PSF is only $\sim$0\farcs5 - 1\farcs0, 
depends on proper aperture corrections 
between the science-field PSF and the photometric standard star's aperture.
The variable AO correction between the
NGS observations of the bright standard stars and the LGS observations,
with faint off-axis tip-tilt stars, of a Galactic star cluster, for instance,
result in different PSFs that cannot easily be matched at radii of
$<$0\farcs5. Sheehy et al.\cite{sheehy06} proposed to model the PSF from the 
science observations directly out to large radii by assuming the 1D power
spectrum of the image can be described entirely from the distribution
of stars, a noise floor, and a PSF constructed from the telescope, 
camera detector, and AO-corrected atmosphere. With this technique they
achieve 4\% photometric accuracy across multiple exposures of the 
same field. However, when compared with HST observations of the
same field they find an RMS error of only 20\%, some of which may be attributed
to intrinsic source variability. Again, improved knowledge of the complete
PSF for crowded stellar fields would provide better aperture corrections
permitting absolute photometric accuracies of $<$5\% for fields photometrically
calibrated only from AO observations, rather than bootstrapping from
lower spatial resolution observations from 2MASS, UKIDSS, or HST. 

\section{Conclusions}
Astrometric and photometric measurements with adaptive optics observations
have achieved accuracies of
\begin{itemize}
\item Relative Astrometric Errors: $<$0.2 mas
\item Relative Photometric Errors: 0.03 - 0.04 mag
\item Absolute Photometric Errors: $<$0.05 mag (however, see discussion 
  in \S\ref{sec:absPhotometry}).
\end{itemize}
The current limitations of such measurements are primarily due to uncertainties
in the PSF. Efforts in PSF-reconstruction from AO telemetry \cite{veran97}, 
modeling the spatial dependence of the PSF from Cn$^2$ profiles 
\cite{britton06}, and better methods for fitting the PSF empirically 
for crowded fields will be needed to improve upon the above accuracies.

\bibliography{jlu,gc_bib,astrometry,clusters}   

\begin{thebibliography}{10}

\bibitem{dupuy09}
{Dupuy}, T.~J., {Liu}, M.~C., and {Ireland}, M.~J., ``{Dynamical Mass of the
  Substellar Benchmark Binary HD 130948BC},'' {\em \apj}~{\bf 692},  729--752
  (Feb. 2009).

\bibitem{konopacky10}
{Konopacky}, Q.~M., {Ghez}, A.~M., {Barman}, T.~S., {Rice}, E.~L., {Bailey},
  J.~I., {White}, R.~J., {McLean}, I.~S., and {Duch{\^e}ne}, G.,
  ``{High-precision Dynamical Masses of Very Low Mass Binaries},'' {\em
  \apj}~{\bf 711},  1087--1122 (Mar. 2010).

\bibitem{ghez08}
{Ghez}, A.~M., {Salim}, S., {Weinberg}, N.~N., {Lu}, J.~R., {Do}, T., {Dunn},
  J.~K., {Matthews}, K., {Morris}, M.~R., {Yelda}, S., {Becklin}, E.~E.,
  {Kremenek}, T., {Milosavljevic}, M., and {Naiman}, J., ``{Measuring Distance
  and Properties of the Milky Way's Central Supermassive Black Hole with
  Stellar Orbits},'' {\em \apj}~{\bf 689},  1044--1062 (Dec. 2008).

\bibitem{gillessen09}
{Gillessen}, S., {Eisenhauer}, F., {Trippe}, S., {Alexander}, T., {Genzel}, R.,
  {Martins}, F., and {Ott}, T., ``{Monitoring Stellar Orbits Around the Massive
  Black Hole in the Galactic Center},'' {\em \apj}~{\bf 692},  1075--1109 (Feb.
  2009).

\bibitem{lu09}
{Lu}, J.~R., {Ghez}, A.~M., {Hornstein}, S.~D., {Morris}, M.~R., {Becklin},
  E.~E., and {Matthews}, K., ``{A Disk of Young Stars at the Galactic Center as
  Determined by Individual Stellar Orbits},'' {\em \apj}~{\bf 690},  1463--1487
  (Jan. 2009).

\bibitem{bartko09}
{Bartko}, H., {Martins}, F., {Fritz}, T.~K., {Genzel}, R., {Levin}, Y.,
  {Perets}, H.~B., {Paumard}, T., {Nayakshin}, S., {Gerhard}, O., {Alexander},
  T., {Dodds-Eden}, K., {Eisenhauer}, F., {Gillessen}, S., {Mascetti}, L.,
  {Ott}, T., {Perrin}, G., {Pfuhl}, O., {Reid}, M.~J., {Rouan}, D.,
  {Sternberg}, A., and {Trippe}, S., ``{Evidence for Warped Disks of Young
  Stars in the Galactic Center},'' {\em \apj}~{\bf 697},  1741--1763 (June
  2009).

\bibitem{stolte08}
{Stolte}, A., {Ghez}, A.~M., {Morris}, M., {Lu}, J.~R., {Brandner}, W., and
  {Matthews}, K., ``{The Proper Motion of the Arches Cluster with Keck
  Laser-Guide Star Adaptive Optics},'' {\em \apj}~{\bf 675},  1278--1292 (Mar.
  2008).

\bibitem{cameronThesis}
{Cameron}, P.~B., {\em {The Formation and Evolution of Neutron Stars:
  Astrometry, Timing, and Transients}}, PhD thesis, Caltech (June 2008).

\bibitem{cameron09}
{Cameron}, P.~B., {Britton}, M.~C., and {Kulkarni}, S.~R., ``{Precision
  Astrometry With Adaptive Optics},'' {\em \aj}~{\bf 137},  83--93 (Jan. 2009).

\bibitem{luThesis}
{Lu}, J.~R., {\em {Orbits and Origins of Young Stars at the Galactic Center}},
  PhD thesis, UCLA (June 2008).

\bibitem{clarkson10}
{Clarkson}, W., {Lu}, J.~R., , {Ghez}, A., {Morris}, M., {McCrady}, N.,
  {Stolte}, A., and {Yelda}, S., ``{A sharper look at the motion of stars in
  the Arches with Keck-LGS adaptive optics.},'' {\em Proceedings of "The
  Galactic Center: A Window on the Nuclear Environment of Disk Galaxies"}
  (2010).

\bibitem{teoste88}
{Teoste}, R., {Daley}, Jr., J.~A., {Capes}, Jr., R.~N., {Alves}, J.~J., and
  {Zimmerman}, M.~D.,  [{\em {Measurements of tilt anisoplanatism at the
  Firepond Facility}}{\nolinebreak\hspace{0.1em}]}, Unknown (Nov. 1988).

\bibitem{gublerTytler98}
{Gubler}, J. and {Tytler}, D., ``{Differential Atmospheric Refraction and
  Limitations on the Relative Astrometric Accuracy of Large Telescopes},'' {\em
  \pasp}~{\bf 110},  738--746 (June 1998).

\bibitem{fritz10}
{Fritz}, T., {Gillessen}, S., {Trippe}, S., {Ott}, T., {Bartko}, H., {Pfuhl},
  O., {Dodds-Eden}, K., {Davies}, R., {Eisenhauer}, F., and {Genzel}, R.,
  ``{What is limiting near-infrared astrometry in the Galactic Centre?},'' {\em
  \mnras}~{\bf 401},  1177--1188 (Jan. 2010).

\bibitem{starfinder}
{Diolaiti}, E., {Bendinelli}, O., {Bonaccini}, D., {Close}, L., {Currie}, D.,
  and {Parmeggiani}, G., ``{Analysis of isoplanatic high resolution stellar
  fields by the StarFinder code},'' {\em \aaps}~{\bf 147},  335--346 (Dec.
  2000).

\bibitem{hornstein07}
{Hornstein}, S.~D., {Matthews}, K., {Ghez}, A.~M., {Lu}, J.~R., {Morris}, M.,
  {Becklin}, E.~E., {Rafelski}, M., and {Baganoff}, F.~K., ``{A Constant
  Spectral Index for Sagittarius A* during Infrared/X-Ray Intensity
  Variations},'' {\em \apj}~{\bf 667},  900--910 (Oct. 2007).

\bibitem{eckart08}
{Eckart}, A., {Baganoff}, F.~K., {Zamaninasab}, M., {Morris}, M.~R.,
  {Sch{\"o}del}, R., {Meyer}, L., {Muzic}, K., {Bautz}, M.~W., {Brandt}, W.~N.,
  {Garmire}, G.~P., {Ricker}, G.~R., {Kunneriath}, D., {Straubmeier}, C.,
  {Duschl}, W., {Dovciak}, M., {Karas}, V., {Markoff}, S., {Najarro}, F.,
  {Mauerhan}, J., {Moultaka}, J., and {Zensus}, A., ``{Polarized NIR and X-ray
  flares from Sagittarius A*},'' {\em \aap}~{\bf 479},  625--639 (Mar. 2008).

\bibitem{do09}
{Do}, T., {Ghez}, A.~M., {Morris}, M.~R., {Yelda}, S., {Meyer}, L., {Lu},
  J.~R., {Hornstein}, S.~D., and {Matthews}, K., ``{A Near-Infrared Variability
  Study of the Galactic Black Hole: A Red Noise Source with NO Detected
  Periodicity},'' {\em \apj}~{\bf 691},  1021--1034 (Feb. 2009).

\bibitem{harayama08}
{Harayama}, Y., {Eisenhauer}, F., and {Martins}, F., ``{The Initial Mass
  Function of the Massive Star-forming Region NGC 3603 from Near-Infrared
  Adaptive Optics Observations},'' {\em \apj}~{\bf 675},  1319--1342 (Mar.
  2008).

\bibitem{doThesis}
{Do}, T., {\em {Physical Processes in the Vicinity of a Supermassive Black
  Hole}}, PhD thesis, UCLA (Aug. 2010).

\bibitem{schoedel10}
{Sch{\"o}del}, R., ``{Accurate photometry with adaptive optics in the presence
  of anisoplanatic effects with a sparsely sampled PSF. The Galactic center as
  an example of a challenging target for accurate AO photometry},'' {\em
  \aap}~{\bf 509},  A58+ (Jan. 2010).

\bibitem{esslingerEdmunds98}
{Esslinger}, O. and {Edmunds}, M.~G., ``{Photometry with adaptive optics: A
  first guide to expected performance},'' {\em \aaps}~{\bf 129},  617--635 (May
  1998).

\bibitem{stolte10}
{Stolte}, A., {Morris}, M., {Ghez}, A., {Do}, T., {Lu}, J., {Wright}, S.,
  {Ballard}, C., {Mills}, E., and {Matthews}, K., ``{Disks in the Arches
  cluster -- survival in a starburst environment},'' {\em ArXiv e-prints}
  (June 2010).

\bibitem{sheehy06}
{Sheehy}, C.~D., {McCrady}, N., and {Graham}, J.~R., ``{Constraining the
  Adaptive Optics Point-Spread Function in Crowded Fields: Measuring
  Photometric Aperture Corrections},'' {\em \apj}~{\bf 647},  1517--1530 (Aug.
  2006).

\bibitem{veran97}
{Veran}, J., {Rigaut}, F., {Maitre}, H., and {Rouan}, D., ``{Estimation of the
  adaptive optics long-exposure point-spread function using control loop
  data.},'' {\em Journal of the Optical Society of America A}~{\bf 14},
  3057--3069 (Nov. 1997).

\bibitem{britton06}
{Britton}, M.~C., ``{The Anisoplanatic Point-Spread Function in Adaptive
  Optics},'' {\em \pasp}~{\bf 118},  885--900 (June 2006).

\end{thebibliography}
\bibliographystyle{spiebib}   

\end{document}